\documentclass[aps,prl,twocolumn]{revtex4}

\usepackage{epsfig}

\newcommand{\be}{\begin{equation}}
\newcommand{\ee}{\end{equation}}
\newcommand{\bea}{\begin{eqnarray}}
\newcommand{\eea}{\end{eqnarray}}

\def\beginwide{
        \end{multicols} \vspace*{-0.5cm} \noindent
        \rule{3.5in}{.1mm}\rule{.1mm}{5mm} \widetext \medskip }
\def\beginwidetop{
        \end{multicols} \vspace*{-0.5cm} \noindent
        \widetext \medskip }
\def\endwide{
        \hspace*{3.35in}~\rule[-5mm]{.1mm}{5mm}\rule{3.5in}{.1mm}
        \begin{multicols}{2} \vspace*{-1.0cm} \noindent }
\def\endwidebottom{
        \begin{multicols}{2} \vspace*{-1.0cm} \noindent }

\begin{document} 
\title{ Probability Distribution of Residence-times of Grains in Sandpile Models}

\author{ Deepak Dhar and Punyabrata Pradhan } 
\address{Department of Theoretical Physics, Tata Institute of Fundamental Research, Homi Bhabha Road, Mumbai-400005, India.}

\begin{abstract} 

We show that the probability distribution of the
residence-times of sand grains in sandpile models, in the scaling limit, 
can be expressed in terms
of the survival probability of a single diffusing particle in a medium
with absorbing boundaries and space-dependent jump rates.  The scaling
function for the probability distribution of residence times is
non-universal, and depends on the probability distribution according to
which grains are added at different sites.  We determine this function
exactly for the 1-dimensional sandpile when grains are added randomly only
at the ends. For sandpiles with grains are added everywhere with equal
probability, in any dimension and of arbitrary shape, we prove that, in
the scaling limit, the probability that the residence time greater than
$t$ is $exp(-t/\bar{M})$, where $\bar{M}$ is the average mass of the pile
in the steady state. We also study finite-size corrections to
this function.

\end{abstract} 

\maketitle



In 1987, Bak, Tang and Wiesenfeld (BTW) proposed sandpile model, as a paradigm of
self-organized critical systems in nature \cite{bak}. Since then, many different
kinds of sandpile models have been studied. These include models with discrete
or continuous variables \cite{zhang}, different toppling rules \cite{knlz}, deterministic or stochastic driving \cite{manna}, with or without local particle conservation \cite{mohanty}, driving mechanisms \cite{fes1,fes2} etc. By now, a fair amount of understanding of the critical steady state and critical exponents of avalanches has been achieved, by the study of exactly solved models,and by numerical simulations. For a review of known results, see \cite{reviews1,reviews2,reviews3,reviews4}.

 However, time-dependent properties of self-organized critical systems
have not been studied as much theoretically so far, in spite of the fact that
an explanation of 1/f noise was one of the main motivations for the initial
proposal of self-organized criticality.  While a power spectrum of
mass-fluctuations of 1/f type has been found in some 1-dimensional models
\cite{ali,maslov}, it appears that in higher dimensional sandpile models,
the behavior is $1/f^2$ \cite{kertesz}. Piles of long-grain rice have
provided a very good experimental realization the basic ideas of
self-organized criticality \cite{kim,frette}.  The probability
distribution of residence times (DRT) of grains was studied experimentally
in the Oslo ricepile experiment, and by simulations \cite{kim,boguna}.

In this paper, we will study DRT in sandpile models. We argue that the
problem of determining DRT can be reduced to that of finding the
probability distribution of hitting time of a single diffusing particle to
the boundary, diffusing in a medium with site-dependent jump rates. In the
scaling limit of large system sizes, DRT becomes a function of a single
scaling variable $t/L^b$, where $t$ is the residence time, $L$ is the 
linear size of the system, and $b$ is some exponent. This function is non-universal, and is a
complicated function of the spatial distribution of  added grains used to 
drive the pile to its steady state. We determine this function explicitly for 1-dimensional
sandpile when grains are added randomly only at the ends. When grains are
added with equal probability everywhere, we prove that the exact scaling
function of the DRT is a simple exponential.  This result is independent
of dimension, and of the shape of the pile.

Let us consider the problem in the simplest setting first: the BTW model 
\cite{bak} 
on a line  of $L$ sites, labeled by integers $1$ to $L$.  At each site 
$i$ we
have a non-negative integer variable $z_i$ called the height of the pile at that site. The
site is stable if $z_i \le 1$. If $z_i \ge 2$, the site is said to be unstable,
and relaxes by toppling. In this process, $z_i$ decreases by $2$, and $z_{i-1}$
and $z_{i+1}$ increase by $1$. Toppling at a boundary site causes the loss of
one sandgrain from the pile. The pile is driven by adding grains at the right
end, at a constant rate of one grain every $P$ time steps. We assume that $P$ is
larger than the duration of the longest avalanche in the system, so that all
avalanches have died before a new grain is added to the system.

This model has an abelian property, and its properties are well-understood
\cite{dhar}. The long-time behavior under the deterministic evolution is
that after an initial transient period, it falls into a cycle of period $L
+ 1$. The stable configurations of the pile that belong to the cycle are
$L$ configurations having all, except one site, with height $1$, and one
configuration with all $z_i = 1$. If we start with the state with all $z_i
=1$, adding a particle at $i=L$ gives a stable configuration with $z_1
=0$. Adding a particle again, we get the recurrent configuration in which
$z_{2} = 0$. For each new added grain, the position of the zero shifts one
step to the right, till after $L$ steps it is at $i = L$. Then adding
another grain, the zero disappears. We choose to say that in this case the
zero is at $i= 0$. The number of toppling to get the next stable
configuration is also periodic with the same period : $ L \rightarrow L -
1 \rightarrow (L-2)  \ldots 1 \rightarrow 0 \rightarrow L $.

If we want to study DRT in this model, we have to mark the grains.  
However, with marked gains, {\it the model is no longer abelian}. This is
because toppling at two adjacent unstable sites in different order no
longer give the same result. For a full specification of the rules
governing the motion of grains in the model, we have to define precisely
in which order the unstable sites are toppled, and how the grains are
transferred under toppling. We choose the parallel update scheme: make a
list of all sites which are unstable at a time $t$, choose at random two
grains from each of these sites ( if there are only two grains, both are
selected), and randomly assign one of them to go the left neighbour, and
the other to the right. All these grains which are to be moved are then
added to their destined sites, at the same time. This constitutes a single
microstep of evolution. Then we construct the new list of unstable sites
for the next micro time-step, and repeat.

The constant time elapsed between two successive additions of grains ($P$ micro-steps) will be called a meso-time step. We measure the residence time in units of meso-steps. We mark all grains by the meso-time step when they were added to the pile. Then, if the grain numbered $T_{in}$ (added at meso-step $T_{in}$) gets out of the system at meso-step $T_{out}$, we will say that its residence time is $T_{out} - T_{in}$.

It is easily seen that the first moment of the DRT is the average mass of the 
pile. Define a variable $\eta(i,j)$ as the indicator function that the 
$i$-th grain is in the pile at the end of time meso-step $j$. Then, 
summation of $\eta(i,j)$ over $j$ gives the residence time of particle 
$i$, and average over $i$ gives the mean residence time. Conversely, sum 
over $i$, we get the mass of the pile at the end of time-step $j$, and average
over $j$ gives the mean mass of the pile \cite{pradhan}. 

From our definition, it follows that the probabilities of different paths
taken by a grain are exactly that of an unbiased random walker on the line. This is
because when a grain moves under toppling, it is equally likely to take a
step to the right, or to the left. So, for example, the average number of
steps a grain takes before it leaves the pile is equal to the average
number of steps a random walker would take from that starting point.  
However, the time between two jumps of the grain is random, and has very
non-trivial correlations with times of previous jumps, and also with jump
times of other particles. This is what makes this problem nontrivial.

To calculate the DRT for the linear chain of $L$ sites, we consider adding a
marked grain at meso-time $T_{0}$. All other grains are unmarked, and
indistinguishable. Then, stable configurations of the pile are $L^2$ in number.
Configuration in which the site $a$ has height $0$, and the marked grain is at
site $b$ will be denoted by ${\cal C}_{a,b}$. All sites other than $a$ and $b$
are occupied by unmarked grains.  For each value of $a$, $1 \le a \le L$, then
there are $L-1$ possible configurations corresponding to different values of
$b$. For the recurrent configuration with all $z_i=1$, we define $a=0$, and in this case there are $L$ possible positions of $b$. Thus there are
in total $L^2$ possible stable configurations of the pile.

Consider a particular configuration ${\cal C}_{a,b}$. Adding another (unmarked)
grain at $i = L$. If $b < a $, then it is easily seen that the wave of toppling
\cite{wave} does not reach the marked grain, and the final configuration is ${\cal
C}_{a+1,b}$.

When $ b > a$, the wave of toppling, started at the right end, reaches the
site $b$ and the site will topple. The marked grain will move one step to the left
or right, with equal probabilities.  If the marked grain moves to the left, it
will move again due to toppling, unless that site has no grains. In this way,
the marked grain can take zero, one or more consecutive steps to the left in one
meso-step. It stops diffusing as soon as it takes a right step or if the
marked grain falls on $a$. We thus see that on adding more grain, if $b > a$,
the final configuration is ${\cal C}_{a +1, b + \Delta b }$, with $\Delta b$
taking values $ 1, 0, -1 \ldots a +1 - b, a - b$ with probabilities $ 2^{-1},
2^{-2}, 2^{-3} \ldots, 2^{a-b+1}, 2^{a-b}$. For $(b-a)$ large, the mean square displacement $\langle (\Delta b)^2 \rangle$ tends to $2$.

It is straight forward to construct the $L^2 \times L^2$ matrix ${\cal W}$ giving the transition probabilities between different configurations. Also, knowing the probabilities of different configurations in the steady state, we can write down the probabilities of different stable configurations just after the marked grain has been added. Then ${\cal W}^t P(t=0)$ gives the probabilities of different configurations at time $t$, and summing over all configurations, we get the
survival probabilities $S(t)$ that the marked grain remains inside the system up to time $t$. In
fact, using the fact that the position $a$ of site with height zero changes by 
$1$ deterministically in time, one can rewrite this problem in terms of (L+1)  matrices ${\cal T}_j$, $j=0,1,2, \ldots L$, where ${\cal T}_j$ is a $(L-1) \times (L-1)$ matrix giving transition
probabilities from the $L-1$ configurations with $a=j$ to the $L-1$
configurations with $a =j +1$. For $a=L+1$, there are $L$ stable configurations
with a marked grain, but the configuration with $b =L$ is transient, and
hence one can work with the $(L-1)$ remaining configurations. 

In an earlier paper, one of us used these to determine DRT exactly numerically, for $L$ up
to 150 \cite{pradhan}.  In the limit of large $L$, it was argued  that 
$Prob(t|L)$ tends to the scaling form

\begin{equation}
Prob( t | L) \sim  \frac{1}{L^3} f( t / L^2).
\end{equation}

Here the function $f(x)$ varies as $x^{-3/2}$ for $x 
\ll 1$, and as $exp(-Kx)$ for $x \gg 1$. We now obtain the exact 
functional form of the scaling function $f$.
 
Consider a marked grain at $b$, at some time $T$, with $b =
\alpha L, 0 < \alpha < 1$, and $L$ large. We consider the change in its
position $\Delta b$ after one cycle ( $ (L + 1)$ mesosteps). Fig.1 shows
the motion of grains in a cycle in one realization. The grain diffuses for
a while, and is stuck when the zero is to the right of the marked grain.
The average fraction of time it moves is $\alpha$.  A grain at $b$ is
hit by $b$ waves of toppling \cite{wave} in this interval. The net displacement $\Delta b$ is sum of displacements due to these waves. Each wave causes a
displacement with mean zero, and variance $2$. Then by central limit
theorem, the net displacement will be distributed normally with variance
given by $ 2 b$. Thus $\Delta b$ is of order $\sqrt { 2 \alpha L}$, and
is much smaller than $L$ when $L$ is very large. Then, for times $t \gg L$, we can average over
the motion in a cycle, and say that if the marked grain is at $i$, it
moves to the left or right neighbor with a rate $ (i/L)$ per unit time. If
$P(i,t)$ is the probability that the marked grain is at $i$ at time $t$,
the evolution equation for $P(i,t)$ for times $t \gg 1$ is

\begin{equation}
\frac{d}{dt} P(i,t) = \frac{i+1}{L} P(i+1,t) +\frac{i-1}{L} P(i-1,t) 
- \frac{2 i}{L} P(i,t).
\label{E2}
\end{equation}

At time $t=0$, we can assume that the marked particle is at $i = L$, 
so that $P(i,t=0) = \delta_{i,L}$.  Integrating this equation, we 
determine the survival probability $S(t) = \sum_{i} P(i,t)$, and then
the DRT is given by 
\begin{equation}
Prob(t|L) = S(t)- S(t+1).
\label{E3}
\end{equation}

We introduce the reduced coordinate $\xi = i/L$, and $\tau = t/L^2$ and consider the Eq.(\ref{E10}) when $L$ is large. In terms of these reduced variables, the evolution equation for the probability density $P(\xi,\tau)$ becomes, in the continuum limit,

\begin{equation} 
\frac{\partial}{\partial \tau} P(\xi,\tau) =
\frac{\partial^2}{\partial \xi^2} [ \xi P(\xi,\tau)].
\end{equation}

We can integrate this equation numerically using the initial condition $P(\xi,t=0)=\delta (\xi-1+1/L)$. The scaling function $f(x)$ is given by, 

\begin{equation}
f(x)= \left[ \frac{d}{d\tau} \int_0^{1} P(\xi,\tau) d\xi \right]_{\tau=x}
\end{equation}

Let $\varphi_j(\xi)$ be solution to the eigenvalue equation corresponding to eigenvalue $\lambda_j$
 
\begin{equation}
\frac{d^2}{d \xi^2} [ \xi \varphi_j (\xi)] = -\lambda_j \varphi_j (\xi)\mbox{,}
\end{equation}
where $\varphi_j(\xi=1) = 0$ corresponding to an absorber being present at 
$i= L+1$. At $\xi=0$, we do not need to assume any 
special condition, as the absorber at $i=0$  is automatically taken 
care of by the fact that rate of jump out of $i$ is $2i/L$, which 
becomes zero at $i=0$.

We look for a solution $\varphi_j(\xi)$ that does not diverge at 
$\xi =0$. Expanding 
$\varphi_j(\xi)$ in a power series, and matching coefficients, we
get 

$$
\varphi_j(\xi) = \sum_{n=0}^{\infty} \frac{ (-\lambda_j \xi)^n}{n! (n+1)!}=  
I_1(2i \sqrt{\lambda_j \xi})/ ({i \sqrt{\lambda_j \xi}}),
$$
where $I_1(x)$ is the modified Bessel function of order 1\cite{bessel}. The eigenvalues
$\lambda_j$ is obtained by imposing the condition $\varphi(\xi=1)=0$.
Thus if the $j$-th zero $I_1(z)$ occurs at $\pm 2i k_j$, then $\lambda_j=k_j ^2$. At large times $t$, $S(t)$ varies as $exp(-Ct/L^2)$, where we
get $C=k_1^2=3.6705$. This value is in good agreement with the value
obtained by extrapolation of estimates obtained by measuring the
coefficients of the exponential determined by exact diagonalization of the
master equation for finite $L$ \cite{pradhan}.

The generalization of these results to $d$ dimensions is straight-forward.  
We consider a d-dimensional sandpile model on a lattice with number of sites $V$. We assume that when a new grain is added,
the site $\vec{x}$ is chosen with probability $r(\vec{x})$. Clearly, the
sum of $r(\vec{x})$ over all sites is $1$.  In the steady state, $n(\vec{x})$, the average number of topplings at $\vec{x}$ per added grain, satisfies the equation ~(using conservation of sand grains)

\begin{equation} 
\nabla ^2 n(\vec{x})=-r(\vec{x}),
\label{E7}
\end{equation}
with $n(\vec{x})=0$ at the boundary. The solution of this equation is

\begin{equation}
n(\vec{x})=\sum_{\vec{x'}} G(\vec{x},\vec{x'}) r(\vec{x'}),
\end{equation}
where $G(\vec{x},\vec{x'})$ is the average number of topplings at 
$\vec{x}$ due to addition of a grain at $\vec{x'}$, and is equal to the
inverse of the toppling matrix $\Delta$ \cite{dhar}.

The important point to realize is that while avalanches in sandpile can
spread quite far, the typical distance traveled by one marked grain in
an avalanche is much smaller than $L$.  In fact,
in many cases, we expect it to be of order 1. During its motion to the
boundary, the marked grain would be involved in a large number of
avalanches. At time-scales much larger than a meso-step, the motion is
diffusive, with the jump-rate out of different sites being  
space-dependent because on the average some parts of the
lattice have more avalanche activity than others.

Consider a grain at site $\vec{x}$ at time $t$.  Let its position be 
$\vec{x}+\Delta\vec{x}$ after $\Delta t$ new 
grains have been added, where $ L^d \gg \Delta t \gg 1$. As the path of 
the grain is an unbiased random walk, we have $\langle (\vec{\Delta x})^2 \rangle = 
s$, where $s$ is the average number of jumps the grain makes in this 
interval. 
Assuming that $|\vec{\Delta x}| \ll L$, and that $n(\vec{x})$ is a slowly 
varying function of $\vec{x}$, we see that  $s$ 
has to be proportional to $n(\vec{x}) \Delta t$, total no of toppling waves during time interval $\Delta t$. Let us say $s=K 
n(\vec{x}) \Delta t$, where $K$ is some constant. Writing $\langle 
(\vec{\Delta x})^2 \rangle= \Gamma (\vec{x}) \Delta t$, we get 

\begin{equation}
\Gamma (\vec{x})=K n(\vec{x}),
\label{E9}
\end{equation}
where the constant $K$ depends on the details of the model.
For large times $t$, the probability-density $P(\vec{x},t)$ satisfies the 
equation

\begin{equation}
\frac{\partial}{\partial t} P(\vec{x},t)=\frac{1}{2}K \nabla^2 [ n(\vec{x}) 
P(\vec{x},t)]
\label{E10}
\end{equation}
with the initial condition is given by 

\begin{equation}
P(\vec{x},t=0)=r(\vec{x}).
\end{equation}

It may be noted that Eq.(\ref{E10}) is not the diffusion equation with
space-dependent diffusion constant $D(\vec{x})$, where the right hand side
would have been of the form $ \nabla( D(\vec{x})  \nabla P(\vec{x}))$. The
net current between two sites depends on the difference in the product $n
P$ at the two sites, and can be non-zero even if $\nabla P$ is zero.

Solving this differential equation, with the condition that $P(\vec{x},t)$ is
zero at the boundary corresponding to the absorbing boundaries, we can determine
$P(\vec{x},t)$ at any time $t$. Integrating over $\vec{x}$ determines the
probability that marked particle remains in the system at time $t$, and the DRT is obtained from the survival probability using Eq.(\ref{E3}).

Consider, as an example, the case of a linear chain with $L$ sites, when we add
particles at each step at either of the two ends with probability $1/2$. In this case, we get $n(x)=\frac{1}{2}$, independent of $x$, and $K=2$. One can then solve the Eq.(\ref{E10}) analytically, and a straightforward calculation gives $S(\tau)=\theta_3(0,\tau)-\theta_3( \pi/2,\tau)$, where $\tau=\pi^2t/2L^2$ and $\theta_3(z,\tau)$ is the Jacobi theta function \cite{theta} defined by
$$
\theta_3(z,\tau)=1+2\sum_{n=1}^{\infty} \exp(-n^2\tau) cos(2nz).
$$

In fig 2, we have plotted the analytically computed survival 
probability $S(t)$ versus $t/L^2$ for this case and compared with the results of simulation for $L=100$ using $10^6$ grains.  We have also shown the result of the numerical integration of Eq.(\ref{E2}) to determine the scaling function in the case where the grains are added only at one end, and compared it with the simulation data
obtained for $L=100$ using $10^6$ grains. Clearly, the agreement is 
excellent. This supports our arguments used to obtain Eq.(\ref{E10}).  

The Eq.(\ref{E10}) is very easy to solve in the special case when sand grains are added randomly at any sites in the system. Clearly here
$r(\vec{x})=1/V$ where $V$ is the number of sites in the lattice. Then $n(\vec{x})$ is a solution to the 
equation

\begin{equation}
\nabla^2 n(\vec{x}) = 1/ V,
\end{equation}
The function $P(x,t)=T(t)$, for all $x$, satisfies Eq.(\ref{E10}) in any dimension if 
$$
\frac{dT(t)}{dt}=-\frac{K}{2V}T(t)
$$

With the initial condition $P(x,t=0)=1/V$, it's easy to see that the full solution is given by
\begin{equation}
P(\vec{x},t) = \frac{1}{V} exp( - K t/ 2V).
\label{E13}
\end{equation}

The probability of survival upto time $t$ is  $V P(\vec{x},t)$, and we 
see that it decays in time as a simple exponential. Using the fact that the mean residence time is the average mass $\bar{M}$ in the pile, we see that $ K /2V = 1/\bar{M}$. This then implies that
\begin{equation}
S(t) = exp( -t/\bar{M}).
\label{E14}
\end{equation}

We note that our derivation depends only on Eq.(\ref{E7}) and Eq.(\ref{E9}). These two equations are valid under the conditions of local conservation of sand grains, transfer of fixed number of grains at each toppling and isotropy.
Thus, the results would be equally applicable to models in which toppling
conditions are different, or the transfer of particles is stochastic,
as in the Manna model.

In Fig 3., we have shown the results of a MC simulation study of the DRT in 
four different models: (a) the 1-dimensional BTW 
model for L=100, (b) the 2-dimensional BTW model defined on a cylinder of 
size $50 
\times 50 $, (c) the 1-dimensional Manna model with $L =100$ 
with rule that if $z_i$ 
exceeds $1$, then 2 particles are transferred, each randomly to one of the 
neighboring sites, and (d) the 2-dimensional Manna model on $50 \times 50$ 
cylinder with two grains transferred at each toppling,  each grain in 
randomly chosen 
direction. The number of particles used in each simulation was $10^6$. We 
plot the probability of survival of the marked grain as a function of 
time $t/\bar{M}$, where $\bar{M}$ was determined from the simulation directly. 
We see 
a very good collapse and agreement with the theoretical 
prediction that $exp(-t/\bar{M})$ for $t \gg 1$.

 For small $t$, the probability distribution is determined by the grains
which are added very near the boundary. Boundary avalanches are not properly
taken care of by our analysis. In particular, it is easy to see that the 
probability of added grain coming out immediatly is nonzero in $d$-dimensions, and varies as $1/L$ for large $L$ (this being the ratio of surface to volume). But Eq.(\ref{E13}) would give this to be ${\cal O}(L^{-d})$. This comes from the fact that near the boundary the height distribution is modified, resulting in the effective $K$ becoming different near the boundary. Eq.(\ref{E13}) is valid for $t \gg 1$. 


In Fig 4., we have plotted the difference $\delta S(t|L)$ between the
survival probability from MC simulation data and the scaling theory prediction [Eq. (\ref{E14})], for a 2-dimensional BTW and the Manna models on an $L \times
L$ square lattice for different values of $L$. We find that the curves for
different $L$ collapse onto each other if $L \delta S(t|L)$ is plotted
versus $t/L^2$, indicating that the correction $\delta S(t|L)$ to the Eq.
(\ref{E14}) has the scaling form

\begin{equation}
\delta S(t|L) \sim \frac{1}{L} g(t/L^2),
\end{equation}
where the correction-to-scaling function $g(x)$ is different for the two 
models.

For the one-dimensional BTW model, with grains added everywhere with equal
probability, we can determine exactly the leading ${\cal O}{(1/L)}$
correction to the scaling solutions Eq.(\ref{E14}). Thus, from the scaling
solution Eq.(\ref{E13}), we get $Prob(T=0 | L) = 1/L + {\cal O}{(L^{-2})}$.
But a straightforward calculation shows that actually $Prob(T=0 | L)
= 2/L + {\cal O}(L^{-2})$. However, $Prob( t | L)$ for $t = 1, 2, \ldots$,
is correctly given to the lowest order by $1/L$. Assuming that the remaining
distribution is a simple exponential, we get

\begin{equation}
S_{1d~ BTW}(t) = \frac{1}{L}\delta_{t,0} + ( 1-  \frac{1}{L}) exp[ -\frac{t}{L}( 1-\frac{a}{L})],
\label{E16}
\end{equation}
where we have used the normalization condition $S(0)=1$, and added a
${\cal{O}}(1/L)$ correction term to the coefficient in the exponential.  
Using the condition that first moment of this distribution is $\bar{M}
=L^2/(L+1)$, we get $a =1/2$. In Fig. 5, have shown the results of a
Monte Carlo simulation of this case for $L =40$ and number of grains$=10^6$, and compared with the theoretical prediction [Eq.(\ref{E16})]. We see that the agreement is very good. 

It is interesting to note that the DRT [Eq.(\ref{E13})] has a simple 
universal form, and does not depend on the critical exponents 
for the distribution of
avalanches, which differ for BTW and Manna models, and also depend on 
dimension in a nontrivial way.
In contrast for the Oslo 
ricepile model, Christensen et al \cite{kim}  and Boguna and Coral 
\cite{boguna} found that the DRT for the 
1-dimension ricepile of size $L$, with $L \gg 1$, does involve nontrivial exponents, and has the form

\begin{equation}
Prob(T | L)    = \frac{1} {L^{\nu}} f( T/ L^{\nu}).
\end{equation} 

The exponent $\nu \simeq 1.3$ is related to 
the roughness of the ricepile surface. The function $f(x)$ takes a 
constant value for $x$ small, and varies as $x^{-b}$ for large $x$, with 
$b \simeq 2.4$. 

 To study such models theoretically, we have to consider sandpile models
where there is a stack at each site into which grains are put in, and there are specific rules about which grains leave the stack. For example, one could assume that incoming grains are put on the top of the stack, and outgoing ones leave from the top. This corresponds to the last-in-first-out rule.  Alternatively, we can choose
the first-in-first-out rule, or choose the outgoing grains
probabilistically, with probability of selection depending on distance of
grain from top of the stack. Clearly, the DRT will have different behavior
for different rules. We hope to address this question in the
future publication.

For the Oslo ricepile model, if we consider only grains those are not
permanently stuck in the pile as constituting ``active mass'' of the
ricepile, all mass above the minimum slope of the pile is active. The configuration with the minimum slope is
recurrent, and will recur infinitely often in the steady state as grains are added to the
pile. Therefore, all the grains added after the pile has reached the
minimum slope have a finite residence-time in the pile.  Then the argument
given earlier in this paper implies that mean active mass of the ricepile
should vary as $L^2$. But the result, obtained in \cite{kim} and \cite{boguna}, shows that it varies as  $L^{1.3}$. The reason for the discrepancy in the estimate of mean mass in \cite{kim, boguna} is presumably due
to the very long residence of the grains which happen to get deeply
embedded, making the estimate of the first moment of the DRT unreliable
from short simulations.

We thank A. Nagar for discussions.
\\
\\

\begin{figure}[h!]
\begin{center}
\leavevmode
\includegraphics[width=8cm,angle=0]{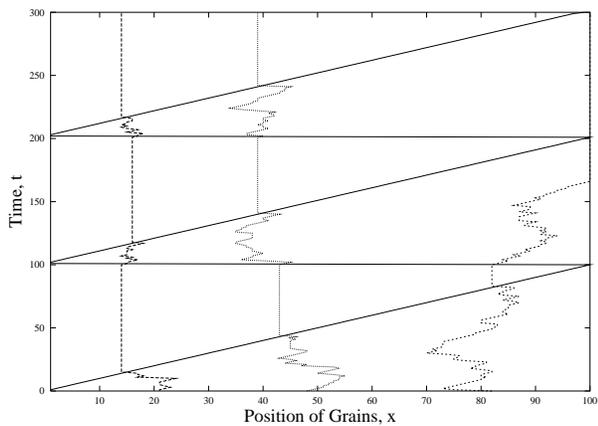}
\caption{\small Motion of three grains starting from $x=20,50,80$ on a one dimensional sandpile of length $L=100$ where sand grains are added only at the right end. }
\end{center}
\end{figure}


\begin{figure}
\begin{center}
\leavevmode
\includegraphics[width=8cm,angle=0]{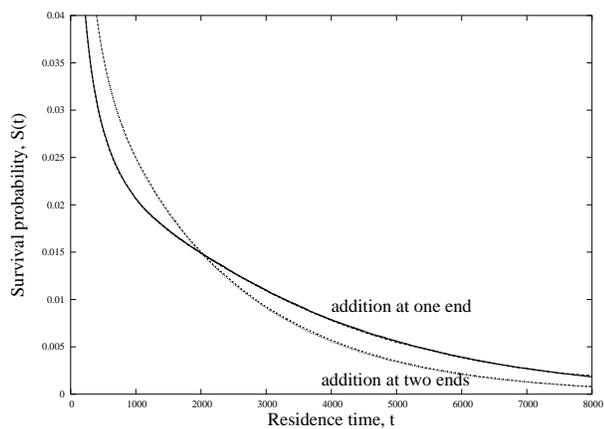}
\caption{\small Survival probability versus residence time $t$ for the 1-dimensional BTW model in two cases with grains added only at one or both ends. The theoretical result(full curve) and the simulation result(dotted line) match perfectly. $L=100$ for both the cases.}
\end{center}
\end{figure}



\begin{figure}
\begin{center}
\leavevmode
\includegraphics[width=8cm,angle=0]{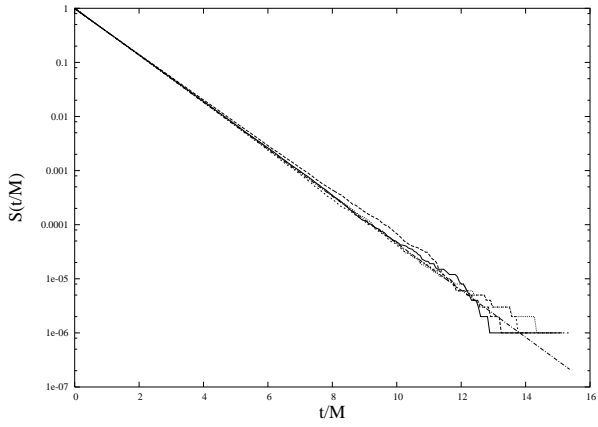}
\caption{\small Semilog plot of the survival probability of the marked grain as a function of scaled residence time $t/M$ for four different cases: (1) one dimensoinal BTW model, (2). one dimensoinal Manna model, (3) two dimensoinal BTW model, (4) two dimensoinal Manna model. We have chosen $L=100$ in one dimension and $70 \times 70$ cylindrical square lattice in two dimensions. }
\end{center}
\end{figure}

\begin{figure}
\begin{center}
\leavevmode
\includegraphics[width=8cm,angle=0]{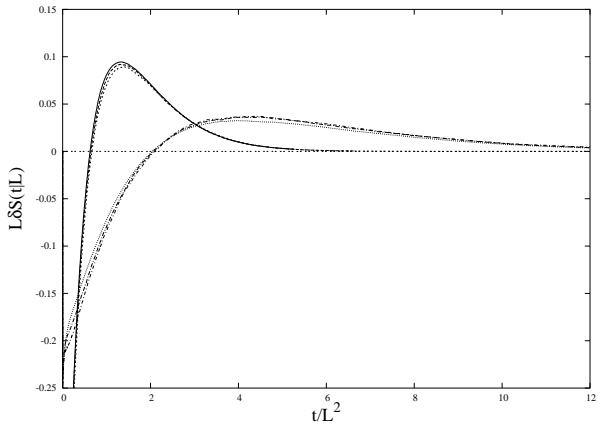}
\caption{\small The scaled correction term $L\delta S(t|L)$ versus $t/L^2$ in $2$-dimensional BTW and Manna model from simulation for three different values of $L=13,20,30$. The curve with higher peak is for the Manna model.}
\end{center}
\end{figure}

\begin{figure}
\begin{center}
\leavevmode
\includegraphics[width=8cm,angle=0]{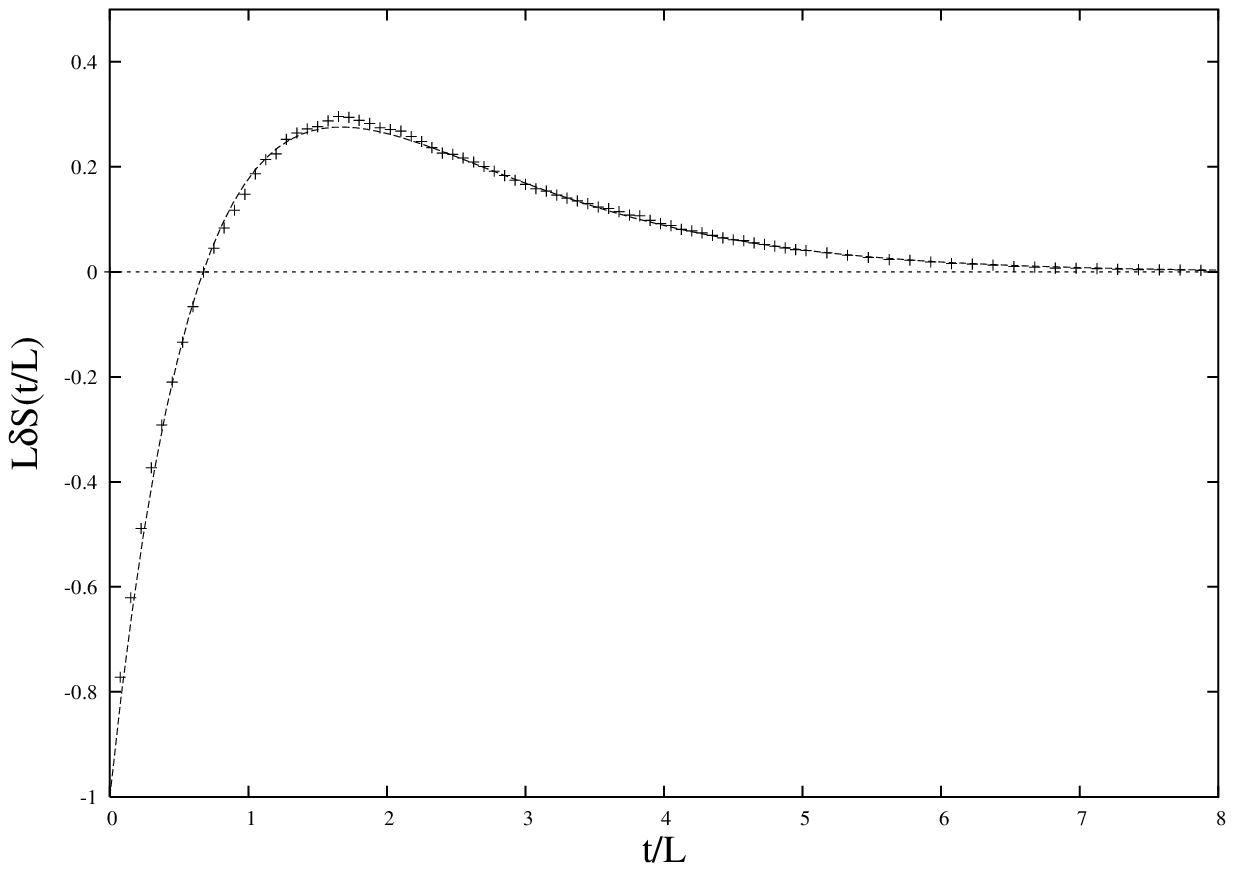}
\caption{\small Finite size correction for the one-dimensional BTW model with grains added everywhere for $L=40$. Plotted is the scaled deviation, $L\delta S(t/L)$, from the simple exponential scaling solution [Eq.(\ref{E14})], of the simulation data and the theory [Eq.(\ref{E16})]. }
\end{center}
\end{figure}



\begin{thebibliography}{99}
\bibitem{bak} P. Bak, C. Tang and K. Wiesenfeld, Phys. Rev. Lett., \newblock{\bf59}, 381, 1988. Phys. Rev. A, \newblock{\bf38}, 364, 1988.

\bibitem{zhang} Y. C. Zhang, Phys. Rev. Lett., \newblock{\bf63}, 470, 1989. 

\bibitem{knlz} L. P. Kadanoff, S. R. Nagel, L. Wu, and S. Zhou, Phys. Rev. A, \newblock{\bf39}, 6524, 1989.
 
\bibitem{manna} S. S. Manna, J. Phys. A: Math. and Gen., \newblock{\bf24}, L363, 1991. 

\bibitem{mohanty} P. K. Mohanty and D. Dhar, Phys. Rev. Lett., \newblock{\bf89}, 104303, 2002.

\bibitem{fes1} A. Chessa, E. Marinari, and A. Vespignani, Phys. Rev. Lett., \newblock{\bf80}, 4217, 1998. 

\bibitem{fes2} R. Dickman, A. M. Munoz, A. Vespignani, and S. Zapperi, Braz. J. Phys., {\bf 30}, 27(2000).

\bibitem{reviews1} H. J. Jensen, {\it Self Organized Criticality} (Cambridge University Press, Cambridge, England, 1998).

\bibitem{reviews2} D. L. Turcotte, Rep. Prog. Phys., {\bf 62}, 1377(1999).

\bibitem{reviews3} E. V. Ivashkevich and V. B. Priezzhev, Physica {\bf A 254}, 97(1998). 

\bibitem{reviews4} D. Dhar, Physica {\bf A 263}, 4(1999).

\bibitem{ali} A. A. Ali, Phys. Rev. {\bf E 52} (1995) R4595.

\bibitem{maslov} S. Maslov, C. Tang, and Y.-C. Zhang, Phys. Rev. Lett. 83, 
2449-2452 (1999).

\bibitem{kertesz} J.  Kertesz and L. B. Kiss, J. Phys. {\bf A 23}, L433 (1990).

\bibitem{kim} K. Christensen, A. Corral, V. Frette, J. Feder, and T. J$\phi$ssang, Phys. Rev. Lett. \newblock{\bf77}, \newblock 107 (1996).

\bibitem{frette} V. Frette, Phys. Rev. Lett. \newblock{\bf70}, \newblock{2762} (1993).

\bibitem{boguna} M. Boguna and A. Corral, Phys. Rev. Lett., \newblock{\bf78}, 4950, 1997.

\bibitem{dhar} D. Dhar, Phys. Rev. Lett., \newblock{\bf64}, 1613, 1990.

\bibitem{pradhan} P. Pradhan and A. Nagar, Paper presented in a Conference (NCNSD-2003) at I.I.T.-Kharagpur, 2003. Cond-mat/0403769.

\bibitem{wave} E. V. Ivashkevich, D. V. Ktitarev, and V. B. Priezzhev, Physica {\bf A 209}, 347(1998).

\bibitem{bessel} M. Abramowitz and I. A. Stegun (Editors), {\it Hand Book of Mathematical Functions} (Dover, New York, 1970) p. 375.

\bibitem{theta} ibid. p. 576 (put $q=\exp(-\tau)$).


\end{thebibliography}
\end{document}